\documentstyle[prl,aps,preprint]{revtex}

\def\be{\begin{eqnarray}}  \def\ee{\end{eqnarray}}  
\def\ba#1{\begin{equation}\begin{array}{#1}}
\def\ea{\end{array}\end{equation}}
\def\lb#1{\label{#1}}    
\def\rf#1{~(\ref{#1})}    \def\ct#1{$^{\cite{#1}}$}
\def\exp#1{e^{#1}}        
\def\xbj{x_{\rm Bj}}          \def\e{{\bf\varepsilon}}
\def\ga{\gamma}		  \def\Ga{\Gamma}
\def\la{\lambda}          
\def\xp{x^{\perp}}  \def\tx{\tilde{x}}

\def\kint#1{\int \frac{\mbox{d}#1}{2\pi}~}
\def\xint#1{\int \mbox{d}#1} 
\def\d{\mbox{d}}
\def\Fd{F^{\stackrel{*}{}}{}}

\def\Journal#1#2#3#4{{#1} {\bf #2}, #3 (#4)}

\def\NPB{{\em Nucl. Phys.} B}
\def\PLB{{\em Phys. Lett.}  B}
\def\PRL{{\em Phys. Rev. Lett.}}
\def\PRD{{\em Phys. Rev.} D}

\def\ZPC{{\em Z. Phys.} C}

\begin{document}
\setcounter{page}{0}
\def\footnoterule{\kern-3pt \hrule width\hsize \kern3pt}
\tighten
\title{    Quark and Gluon Orbital Angular \\ 
	Momentum and Spin in Hard Processes\thanks
{This work is supported in part by funds provided by the U.S.
Department of Energy (D.O.E.) under cooperative 
research agreement \#DF-FC02-94ER40818.}}

\author{S.~V.~Bashinsky and R.~L.~Jaffe\footnote[0]{Email addresses: 
{\tt sergei@mit.edu}, {\tt jaffe@mit.edu}; ~phone: (617)-253-4858; 
~fax: (617)-253-8674.}}

\address{{~}\\Center for Theoretical Physics \\
Laboratory for Nuclear Science \\
and Department of Physics \\
Massachusetts Institute of Technology \\
Cambridge, Massachusetts 02139 \\
{~}}

\date{MIT-CTP-2733,~hep-ph/9804397. {~~~~~} April 1998}
\maketitle

\thispagestyle{empty}

\begin{abstract}
We suggest a method of constructing gauge invariant quark and gluon 
distributions that describe an abstract QCD observable and
apply this method to analyze angular momentum
of a hadron. In addition to the known quark and gluon polarized structure 
functions, we obtain gauge invariant distributions for quark and gluon 
orbital angular momenta, and consider some basic properties of these 
distributions and their moments.
\end{abstract}

\newpage

\section{Introduction}

Since the initial realization that quark spin accounts 
for only a small fraction 
of the nucleon spin\ct{EMC}, theorists have struggled to 
define\ct{Ratcl} and suggest ways to measure the other 
components of the angular momentum in QCD. 
At first even the meaning of the quark spin contribution, $\Sigma$, to the nucleon
spin was the subject of considerable debate. Now it is gradually agreed that 
different interpretations of $\Sigma$ reduce to renormalization scheme dependence 
made particularly irksome by the role of the axial anomaly in this channel. 
The gluon spin fraction, $\Delta G$, seems to be well defined in the context 
of renormalization group improved parton models\ct{C&S}, 
where the polarized gluon structure
function $\Delta g(\xbj,Q^2)$ has the meaning\ct{dG,BJdG} 
of a scale dependent gluon spin 
distribution over the Bjorken variable $\xbj$\footnote{
As usual, $\xbj\equiv Q^2/2P{\cdot}q$ where $P$ is the nucleon
4-momentum, $q$ is the momentum transfer to the nucleon, and $Q^2=-q^2$. 
} so that
\be
\Delta G(Q^2) = \xint{\xbj}~\Delta g(\xbj,Q^2)~.
\lb{dGdef}
\ee
Of course, $\Delta g$ and $\Delta G$ are physical, gauge invariant 
quantities and experiments are being pursued to measure them. 
Nevertheless, there exists no local gauge invariant operator that could 
represent the space density of the gluon spin, a fact that has complicated
the description of the gluon spin fraction.

One may expect to incorporate the existing quark and gluon spin fractions 
into the nucleon angular momentum sum rule\ct{J&M}
\be
\frac{1}{2}~\Sigma + L_q + \Delta G + L_g = \frac{1}{2}~.
\lb{sumrule}
\ee
However, the situation with respect to the quark and gluon orbital angular momenta
$L_q$ and $L_g$ is less satisfactory. There are no generally agreed upon 
definitions of $L_q$ and $L_g$, and no associated {\it gauge invariant}
definitions of their parton distributions\ct{newJi}, 
$f_{L_q}(\xbj,Q^2)$ and $f_{L_g}(\xbj,Q^2)$.
Some authors\ct{JTH,H&S,H&K} use $L_q$ and $L_g$ given by the gauge variant 
decomposition of ref.\cite{J&M}, which appears in eq.\rf{J4} of our paper, 
in the gauge $A^+{=}~0$,
but alternative gauge fixing methods\ct{newJi,Ji} might also be considered.
Besides, $A^+{=}~0$ does not completely fix the gauge,
and  $L_q$ or $L_g$ obtained this way are {\it not} separately
invariant under the residual gauge transformation, 
that is not permissible for physical objects.

We give a positive resolution to these questions in Sec.~3 of this paper,
where we obtain scale dependent but otherwise unambiguous distribution
functions for every term in eq.\rf{sumrule} in a single framework.
This will be contrasted with the conventional gauge invariant operator
description\ct{Ji} of angular momentum in a gauge theory, where the 
decomposition\rf{sumrule}
and the polarized gluon distribution $\Delta g$ are obscure and might even
seem unphysical.
But first we find it useful to develop a general approach to the construction 
of parton distributions associated with an arbitrary physical
observable that may characterize quark and gluon fields. This approach is presented 
in the Sec.~2. The premises and results of Secs.~2 and~3 are summarized in the
conclusion.

\section{Structure functions for a physical observable}

We want to describe some additive physical quantity $\Ga$ relative to a deep 
inelastic process, such as inclusive leptoproduction, where the Bjorken variable
$\xbj\equiv{Q^2}/{2P{\cdot}q}$ is fixed
by the kinematics of the experiment (scattering angles and energies)
and the information about the hadron state is limited to forward hadron matrix
elements of certain operators.
In a naive free parton model and the Bjorken limit $q^-\to\infty$
\be
\xbj=\frac{p^+}{P^+},
\ee
where $p^+$ and $P^+$ are the ``$+$'' components of the struck parton and
the target momenta correspondingly. Therefore a similar experiment 
sensitive to $\Ga$ would directly provide ``the density of 
$\Ga$'' carried by quark and gluon partons with the
momentum $p^+=\xbj P^+$, rather than $\Ga$ itself.
With this in mind, we construct gauge and boost invariant expressions
for such densities that we call quark and gluon $\Ga$ distributions,
$f_{\Ga^q}(\xbj)$ and $f_{\Ga^g}(\xbj)$.
The observable $\Ga$ can be defined by any hermitian
operator constructed from quark and gluon fields or as a generator
of some fields transformation. We do not associate
$\Ga$ with a specific experiment measuring it. Therefore, in this paper 
we will not give a universal answer to the question: In what experiment 
one should measure a general observable, but we show what knowledge about 
$\Ga$ can {\it in principle} be obtained in a deep inelastic process.

In order to construct the parton distribution of $\Ga$ in QCD
we should make clear what a quark or gluon 
parton is in an interacting theory. The subtlety here is in the issue
of gauge invariance: a pure quark field in one gauge is a superposition 
of quarks and gluons in another. 
Different ways of gluon field gauge fixing predetermine different decomposition
of the coupled quark-gluon fields into quark and gluon degrees of freedom.
Similarly, one can generalize a gauge variant nonlocal operator, obtained,
for example, by Fock space decompositions of free fields,
to more than one gauge invariant expressions, raising the
problem of deciding which is the ``true'' one. 
We choose to {\it define} quark and 
gluon partons in fully interacting QCD such that their distributions
in $\xbj$ are given by the unambiguous experimentally accessible structure 
functions\ct{C&S} $q(\xbj)$ and $g(\xbj)$. We find that this definition also
reproduces the measurable polarized structure functions $\Delta q(\xbj)$ and 
$\Delta g(\xbj)$.

The required definition of QCD partons is also very appealing
from mathematical point of view.
The ``gauge covariant'' quanta of quark and gluon fields with a definite 
momentum $p^{+}=\xbj P^+$ are the eigenstates of the generators of the
covariant translations in the $x^-$ direction:
\ba{llll}
T^q_-(a^-)~: & \psi(x) & \to & U(x,x+a^-)~\psi(x+a^-)~, \\
T^g_-(a^-)~: & D_{\la}(x) & \to & U(x,x+a^-)
		~D_{\la}(x+a^-)~U(x+a^-,x)~,
\lb{tfin}
\ea
where $x+a^-$ is the shorthand for $(x^+,x^-{+}~a^-,x^1,x^2)$,
$U(x,x{+}a^-)$ and $U(x{+}a^-,x)$  are the Wilson lines 
along the straight path between $x$ and $x+a^-$,
\be
U(x,x+a^-)\equiv 
  P\mbox{exp}[{ig\int^{x}_{x+a^-}d\xi^-A^{+}(\xi)}]~
\lb{Udef}
\ee
with a similar equation for $U(x{+}a^-,x)=U^{\dagger}(x,x{+}a^-)$, 
and $D_{\la}(x)\equiv\partial_{\la}-igA_{\la}(x)$, so that the second
line in eq.\rf{tfin} defines a transformation of the gluon
field $A_{\la}(x)$. The transformation $T^q_-$ commutes with
$T^g_-$ and each of them forms a one-parameter Lie group:
\ba{c}
[T^q_-(a^-),T^g_-(b^-)]=0~,\\
T^q_-(a^-_2)T^q_-(a^-_1)=T^q_-(a^-_1+a^-_2)~,\\
T^g_-(b^-_2)T^g_-(b^-_1)=T^g_-(b^-_1+b^-_2)~.
\ea
These properties are easily established from the fact that
$T^g_-(a^-)$ leaves $A^+(x)$, and consequently the 
Wilson line\rf{Udef}, invariant, as evident from 
the infinitesimal form of the transformations\rf{tfin}
\ba{lll}
\delta^q_- \psi = a^-D_-\psi~,&&\\
\delta^g_- A_{\la} = a^-F_{-\la}~&\Rightarrow&~
\delta^g_- A^+ = 0~.
\lb{tinf}
\ea
Obviously, the transformations $T^q_-$ and $T^g_-$, as defined by
eq.\rf{tfin} or eq.\rf{tinf}, commute with gauge transformations.

Action of both $T^q_-(a^-)$ on the quark field and $T^g_-(a^-)$,
with the {\it same} parameter~$a^-$, on the gluon field yields ordinary 
translation of these fields by $a^-$ plus an overall gauge 
transformation. This additional gauge transformation does not affect any 
gauge invariant observable, therefore the gauge covariant 
translation of {\it every} field in 
the system is physically equivalent to their usual translation.
This is not true, however, when one considers the 
transformations\rf{tfin} for quarks and gluons individually. 
For example, application of $T^q_-(a^-)$ to the quark field 
$\psi(x)$ without changing the gauge field $A_{\la}(x)$
gives a physically different field configuration from what would 
be obtained by the usual translation $\psi(x)\to\psi(x{+}a^-)$,
because a change of the $\psi(x)$ phase without a compensatory
change in $A_{\la}(x)$ is not a symmetry of QCD.

The quark and gluon fields $\psi(x)$ and $A_{\la}(x)$ can be 
decomposed into a sum of one-dimensional irreducible representations 
of the groups $T^q_-$ and $T^g_-$, which are the eigenstates of the
corresponding generators. 
We will see that this decomposition specifies the desired
parton decomposition in an interacting theory. 
The irreducible representations of $T^q_-$ and 
$T^g_-$ become trivial in the light-cone
gauge $A^{+}{=}~0$ where these transformations
coincide with ordinary translations.
Let us denote fields in this gauge by the 
superscript `$LC$'. Then
\ba{l}
\delta^q_- \psi^{LC} = a^-\partial_-\psi^{LC}~,\\
\delta^g_- A^{LC}_{\la} = a^-\partial_- A^{LC}_{\la}~.
\lb{tph}
\ea
and the decomposition that we are seeking for is simply
\be
\psi^{LC}(x)=\kint{k^+} \exp{-ik^+ x^-} 
	\tilde{\psi}(k^+,\tx)~, \lb{pd1}\\
A^{LC}_{\la}(x)=\kint{k^+} \exp{-ik^+ x^-}
	\tilde{A}_{\la}(k^+,\tx)~
\lb{pd2}
\ee
where
\be
\tx \equiv (x^+,x^1,x^2)~.
\ee
Since $T^q_-$ and $T^g_-$ commute with gauge transformation,
their action on fields in an arbitrary gauge
\be
\psi(x) = \exp{i\alpha(x)}\psi^{LC}(x)~~,~~
A_{\la}(x) = \exp{i\alpha(x)}(A^{LC}_{\la}(x)+
	\frac{i}{g}~\partial_{\la})\exp{-i\alpha(x)}~,
\lb{gt}
\ee
amounts to translation of the ``physical'' components
$\psi^{LC}$ and $A^{LC}_{\la}$ according to eq.\rf{tph}
with the unphysical phase $\alpha(x)$ left invariant:
\ba{llll}
T^q_-(a^-),T^g_-(a^-)~: & \alpha(x) & \to & \alpha(x)~.
\ea
Therefore, the ``parton decomposition'' of fields in an 
arbitrary gauge is simply obtained by a gauge transformation
of eqs.\rf{pd1} and \rf{pd2}. These equations can be viewed as 
transition from $x$ to $(k^+,\tx)$ representation which is more suitable 
for deep inelastic scattering, but this transition is done in a way
respecting gauge invariance and coincides with brute-force Fourier
transformation along $x^-$ direction {\it only} in the specific gauge
$A^+{=}~0$.

Our use of the adjective ``physical'' for fields in the gauge $A^+{=}~0$ 
has been somewhat premature.
There remain gauge transformations, for which
the phase $\alpha(x)$ in eq.\rf{gt} does not depend on 
$x^-$, that preserve the condition $A^+{=}~A_-{=}~0$.
All true physical observables must be invariant
under this {\it residual} gauge symmetry as well. 
The fields $\tilde{\psi}(k^+,\tx)$ and 
$\tilde{A}_{\la}(k^+,\tx)$ defined by eqs.\rf{pd1} 
and\rf{pd2} transform under it as
\ba{lll}
\tilde{\psi}(k^+,\tx) & \to & \exp{i\alpha(\tx)}
		\tilde{\psi}(k^+,\tx)~, \\
\tilde{A}_{\la}(k^+,\tx) & \to & \exp{i\alpha(\tx)}
	(\tilde{A}_{\la}(k^+,\tx)+\frac{2\pi i\delta(k^+)}{g}
	~\partial_{\la})\exp{-i\alpha(\tx)}~.
\lb{rgt}
\ea
Notice that the inhomogeneous term in the residual gauge 
transformation of $\tilde{A}_{\la}$ appears only at
$k^+{=}~0$. We find it useful to split the 
field $\tilde{A}_{\la}(k^+,\tx)$ into a part $\tilde{G}_{\la}(k^+,\tx)$,
transforming homogeneously, and
a gauge field of the residual gauge group, ${\cal A}_{\la}(\tx)$:
\be
\tilde{A}_{\la}(k^+,\tx)\equiv 2\pi\delta(k^+)
	{\cal A}_{\la}(\tx)+\tilde{G}_{\la}(k^+,\tx)~,
\lb{AGdef}
\ee
so that under a residual gauge transformation
\be
{\cal A}_{\la}(\tx) \to \exp{i\alpha(\tx)}
	({\cal A}_{\la}(\tx)+\frac{i}{g}~
	\partial_{\la})\exp{-i\alpha(\tx)}~,~~
\tilde{G}_{\la}(k^+,\tx) \to \exp{i\alpha(\tx)}
	\tilde{G}_{\la}(k^+,\tx)\exp{-i\alpha(\tx)}~.
\lb{rgt1}
\ee
The decomposition\rf{AGdef} becomes unambiguous if we 
require
\be
\left.\tilde{G}_{\la}(k^+,\tx)\right|_{k^+=~0} \equiv 0~.
\ee
Then combining eqs.\rf{pd2} and\rf{AGdef} one obtains
\be
{\cal A}_{\la}(\tx)=\frac{1}{\xint{x^-}}
	\xint{x^-}A^{LC}_{\la}(x)~.
\lb{calA}
\ee
The fields $\tilde{\psi}(k^+,\tx)$, $\tilde{G}_{\la}(k^+,\tx)$,
and the residual gauge field ${\cal A}_{\la}(\tx)$ 
provide the basis in the field space that we use to construct
gauge invariant quark and gluon distributions for 
physical observables.

We associate a physical quantity $\Ga$ with the field transformation
generated by the quantum operator corresponding to $\Ga$. 
We do not insist on the invariance of the QCD action under this field 
transformation. If it is a symmetry of the theory then there exists 
a corresponding conserved Noether current, and
its charge, which will be the operator of $\Ga$, 
commutes with the Hamiltonian of the system.
If it is not, the quantum expectation value of $\Ga$ 
for a particular state is still a well defined quantity,
and spin or orbital angular momentum, which are not separately 
conserved in a relativistic system, provide such examples.

Of the possible choices of $\Ga$, only those that commute with
the parton-defining transformations $T^q_-$ and $T^g_-$
can be interpreted as physical properties of quarks and gluons
distributed in $\xbj$ -- that is, as ``parton distributions of~$\Ga$''.
Thus we restrict ourselves to $\Ga$'s that commute with 
$T^q_-$ and $T^g_-$~:
\be
[\delta^q_-,\delta_{\Ga}]=[\delta^g_-,\delta_{\Ga}]=0~.
\lb{[tga]}
\ee
This allows complete description of $\Ga$ in a hadron in terms of its
distributions over partons, $f_{\Ga^{q,g}}(\xbj)$.

These distributions, {\it i.e.} parton densities of $\Ga$,
are easily constructed using the following trick:
We apply $\Ga$ independently to each subspace of fields
with a definite~$k^+$.
Remembering that by eq.\rf{[tga]}
$\Ga$ can be diagonalized in the basis of $T^q_-$ and $T^g_-$
eigenfunctions, we consider the set of independent transformations 
labeled by a continuous parameter~$\xbj$:
\ba{c}
\delta^{(\xbj)}_{\Ga^{q}} \tilde{\psi}(k^+) = 
        i\epsilon~\Ga^{q} \tilde{\psi}(p^+) 
        \left.\delta(\frac{k^+ - p^+}{P^+})~\right|_{p^+ = \xbj P^+}~~~\\
\delta^{(\xbj)}_{\Ga^{g}} \tilde{G}_{\la}(k^+) =
        i\epsilon~\Ga^{g\chi}_{\la} \tilde{G}_{\chi}(p^+) 
        \left.\delta(\frac{k^+ - p^+}{P^+})~\right|_{p^+ = \xbj P^+}~,
        \lb{gp}\\
\ea
and
\ba{c}
\delta^{(\xbj)}_{\Ga^{q,g}}{\cal A}_{\la}=0~\lb{ga},
\ea
where $\Ga^{q}$ and $\Ga^{g}$ are the generators of $\delta_{\Ga}$
restricted to the fields $\tilde{\psi}(k^+,\tx)$ and 
$\tilde{G}_{\la}(k^+,\tx)$ with $k^+=\xbj P^+$. 
The generators $\Ga^{q}$ and $\Ga^{g}$ can be operators acting on spinor, 
Lorentz, color,
{\it etc.} indices as well as on the coordinate $\xp{\equiv}(x^1,x^2)$.
For simplicity we slightly limit our choice of~$\Ga$ further, 
assuming that 
$\Ga^{q}$ and $\Ga^{g}$ do not depend on the parameter~$q^+$ and
do not involve the coordinate~$x^+$ or derivatives with respect to it.
This is not absolutely necessary but it simplifies 
the following equations. We will not need such complications for any of the 
quantities that are required for angular momentum description.

The generators of the field transformations defined by eqs.\rf{gp} and\rf{ga}
are associated with a physical observable 
which is their normalized expectation
in a hadron state~$|P\rangle$. 
These observables will be shown to be the $\Ga$ 
parton distributions that we are looking for.
Applying Fourier transform to eqs.\rf{gp} and\rf{ga}, one finds the 
action of $\delta^{(\xbj)}_{\Ga}$ on the conventional fields $\psi^{LC}(x)$ and 
$A^{LC}_{\la}(x)$~:
\be
\delta^{(\xbj)}_{\Ga^{q}}\psi^{LC}(x)=
	\frac{i\epsilon P^+}{2\pi}\xint{\xi^-}\exp{i\xbj P^+\xi^-}
	\Ga^{q}\psi^{LC}(x+\xi^-)~,~~~~~~~~~~~~\lb{dpsi} \\
\delta^{(\xbj)}_{\Ga^{g}}A^{LC}_{\la}(x)=
	\frac{i\epsilon P^+}{2\pi}\xint{\xi^-}\exp{i\xbj P^+\xi^-}
	\Ga^{g\chi}_{\la}\left(A^{LC}_{\chi}(x+\xi^-)-
	{\cal A}_{\chi}(\tx)\right)~.\lb{dga}
\ee
These transformations are generated by the {\it light-front} charges
\be
Q^{(+)}\equiv\int\d^2\xp\xint{x^-}J^+(x)
\ee
of the corresponding canonical currents\footnote{
By the canonical current for a field transformation
$\delta_{\Ga}\Phi=i\epsilon\Ga(\Phi)$ we mean
$$
J^{\mu}_{\Ga}\equiv \sum_{\stackrel{\mbox{\scriptsize all}}
{\mbox {\scriptsize fields}}}
\frac{\partial{\cal L}}{\partial(\partial_{\mu}\Phi)}(-i)\Ga(\Phi)~.
$$ 
}~: 
\be
J^{\mu}_{\Ga^q}(x;\xbj)=
	\frac{P^+}{2\pi}\xint{\xi^-}\exp{i\xbj P^+\xi^-}
        \bar\psi(x)\ga^{\mu}
	\Ga^{q}\psi(x+\xi^-)~,
\lb{jg1}~~~~~~~~~~~~\\
J^{\mu}_{\Ga^g}(x;\xbj)=
	\frac{i P^+}{2\pi}\xint{\xi^-}\exp{i\xbj P^+\xi^-}
	F^{\mu\la}(x)\Ga^{g\chi}_{\la}
	\left[A_{\chi}(x+\xi^-)-
	{\cal A}_{\chi}(\tx)\right]~.
\lb{jg2}
\ee
It is necessary to use the light-front charges instead of the usual
$Q=\int\d^3 x~J^0(x)$ because the 
transformations\rf{dpsi}--(\ref{dga}) are non-local in~$x^-$.
We can imagine boosting ourselves to the infinite momentum 
frame, $K'$, where the target hadron has an infinitely large momentum in the 
$\hat{x}'_3$ direction: 
\be
P'=(E'_P,0,0,P'^3)~~\mbox{with}~P'^3\to +\infty~.
\ee
In this limit the integration plane in 
\be
Q'=\int\d^3 x'J'^0(x')
\ee
``approaches'' the $x^-$ axis. On the other hand,
at least for fields confined in a massive hadron,
\be
\lim_{P'^3\to +\infty}Q' = Q^{(+)}~,
\ee 
therefore $Q^{(+)}$ is the correct generator.
One may also check this using the light-cone formulation of a gauge 
theory\ct{Kogan}. The light-front charge $Q^{(+)}$ is invariant under
boosts along $\hat{x} _3$ and can be evaluated in any frame in which the hadron 
has zero transverse momentum, $P^1{=}P^2{=}~0$, including the rest frame.
Notice that if the currents\rf{jg1},\rf{jg2} were conserved,
their usual charges, $Q$, would be frame independent and equal to the light-front
ones, $Q^{(+)}$. 

Now we define the quark or gluon {\it distribution functions} 
for the physical observable $\Ga$ in a covariantly normalized
hadron state $|P\rangle$, 
$\langle P'|P\rangle=2E_P(2\pi)^3\delta^{(3)}({\bf P}'-{\bf P})$,
as the normalized expectation
\be
f_{\Ga^{q,g}}(\xbj)\equiv
\frac{1}{\langle P|P\rangle}~\langle P|~Q^{(+)}_{\Ga^{q,g}}(\xbj)~|P\rangle=
~~~~~~~~~~~~~~~~~~~~~~~~~~~~~\nonumber\\
=\frac{1}{2P^+(\int\d^2\xp\xint{x^-})}~\langle P|
\int\d^2\xp\xint{x^-}
J^{+}_{\Ga^{q,g}}(x;\xbj)~|P\rangle~.
\lb{fdef}
\ee
The distributions are normalized so that $f_{\Ga^{q,g}}(\xbj)=
\Ga_{P}\delta(1{-}\xbj)$
in a free parton model with $|P\rangle$ replaced by a quark or gluon parton which 
is an eigenstate of the transformation\rf{gp} with the eigenvalue $\Ga_{P}$.
With such normalization, the sum of the first moments of the quark and gluon 
$\Ga$ distributions gives exactly the expectation value of the observable $\Ga$ for
the hadron state $|P\rangle$ in the infinite momentum frame:
\be
\int_{-1}^{1}\d\xbj\left(f_{\Ga^q}+f_{\Ga^g}\right) =
\frac{\langle P|Q^{(+)}_{\Ga}|P\rangle}{\langle P|P\rangle}=
\lim_{P'^3\to\infty}\frac{\langle P'|Q'_{\Ga}|P'\rangle}
{\langle P'|P'\rangle}~.
\lb{1mom}
\ee

The objects defined by eq.\rf{fdef} are Lorentz boost invariant,
provided $\Ga^{q}$ and $\Ga^{g}$ transform as implied by 
their Lorentz and spinor indexes. They are invariant
under residual gauge transformation if $\Ga^{q}$ and $\Ga^{g}$
transform homogeneously as:
\be
\Ga^{q} \to \exp{i\alpha(\tx)} \Ga^{q} \exp{-i\alpha(\tx)}~,~~
\Ga^{g} \to \exp{i\alpha(\tx)} \Ga^{g} \exp{-i\alpha(\tx)}~.
\lb{Garest}
\ee
Some examples are $\Ga^{q}=1,\ga^{\mu},
\partial_{\perp}{-}~ig{\cal A}_{\perp}$ and  
$\Ga^{g\chi}_{\la}=\delta^{\chi}_{\la}$;
we will have more examples in Sec.~III~. When $f_{\Ga^q}(\xbj)$
and $f_{\Ga^g}(\xbj)$ are invariant under the residual gauge 
group, they are by construction the same in an {\it arbitrary} 
gauge. An explicitly gauge covariant expression 
for the distributions is straightforwardly obtained by inserting the 
Wilson lines $U(x,x{+}\xi^-)$\lb{gi} into eqs.\rf{jg1} and\rf{jg2} as
$\bar\psi{\dots}\psi\to\bar\psi U{\dots}\psi$,
$F{\dots}A\to\mbox{Tr}[FU{\dots}AU^{\dagger}]$.
Even though the currents in eq.\rf{fdef} are not necessarily 
conserved and the charge operator depends on $x^+$, 
the diagonal matrix element between 4-momentum eigenstates 
$|P\rangle$ are $x^+$ independent. 

The composite operators in eqs.\rf{jg1} and\rf{jg2} have ultraviolet 
divergences\ct{UVD_C_p447}, which arise in perturbation 
theory as infinite loop integrals over ${k}_{\perp}$ in the products of two 
fields at the same ${\xp}$. These divergencies are regulated by
a cutoff at ${k}_{\perp}^2\sim Q^2\equiv-q^2$
leading to factors like $\alpha(\mu^2)\log{(Q^2/\mu^2)}$. 
In practice they are handled with standard
methods of operator renormalization. Renormalization of objects 
similar to eq.\rf{fdef} is described in the work\ct{C&S} by Collins 
and Soper.

For our purposes we take the state $|P\rangle$ in the definition of 
distribution
functions to be the hadron helicity eigenstate with the maximal possible helicity,
{\it e.g.} $+\frac{1}{2}$ for a nucleon. 
Of course, for spin independent structure functions 
one could equally well average over helicities. 
For some problems, for example transversity 
distributions\ct{trans}, one must consider 
off-diagonal matrix elements between states with different 
helicities but the same 4-momentum $P$.

Notice that the very last term, ${\cal A}_{\la}(\tx)$,
in eq.\rf{jg2} does not contribute to the $\Ga$ distributions.
Indeed, it produces the following term in $f_{\Ga^g}(\xbj)$~:
\be
\left.f_{\Ga^g}(\xbj)\right|_{\cal A}=
-\frac{i\delta(\xbj)}{2P^+}~
\frac{\langle P|\int\d^2\xp\xint{x^-}\xint{y^-}~
  F^{+\la}(x)\Ga^{g\chi}_{\la}
  A_{\chi}(y^-,\tx)~|P\rangle}
 {~\int\d^2\xp\left(\xint{x^-}\right)^2}~,
\lb{fa}
\ee
where eq.\rf{calA} was used. Because of the $\xbj$
delta-function, this term trivially vanishes for $\xbj\not=0$. 
One might worry that its value at $\xbj=0$ could 
affect the first moment of $f_{\Ga^g}(\xbj)$. In fact, it does not: In the 
$A^+{=}~0$ gauge $F^{+\la}(x)=\partial_-A^{\la}(x)$ and 
$A^{\la}(x)$ remains finite when $x^-{\to}\pm\infty$, thus
the expression\rf{fa} vanishes as an infinite volume average of 
the derivative of a bounded function\footnote{
In this discussion we treat gluon fields as if they where
classical. Since eq.\rf{fa} deals with physical expectation
values, the conclusion $\left.\Delta f_{\Ga^g}(\xbj)\right|_{\cal A}\equiv 0$ 
should remain valid for the quantum case.}.

The familiar structure 
functions $q(\xbj)$, $\Delta q(\xbj)$, $g(\xbj)$, $\Delta g(\xbj)$ 
are instantly reproduced from eq.\rf{fdef}. 
Indeed, the quark number and chirality are counted by $\Ga^q=1$
and $\Ga^q=\ga^5$ correspondingly. Translating the current matrix elements 
to the origin and canceling out the volume factors in the numerator and 
denominator of eq.\rf{fdef} we find
\be
\Ga^q=1 ~~~ &\Rightarrow& ~~~ f_{q}(\xbj)=
\frac{1}{2\pi\sqrt{2}}\xint{\xi^-}\exp{i\xbj P^+\xi^-}
        \langle P|\psi_{+}^{\dagger}(0)
	\psi_{+}(\xi^-)~|P\rangle~,~ \\
\Ga^q=\ga^5 ~~ &\Rightarrow& ~~ f_{\Delta q}(\xbj)=
\frac{1}{2\pi\sqrt{2}}\xint{\xi^-}\exp{i\xbj P^+\xi^-}
        \langle P|\psi_{+}^{\dagger}(0)\ga^5
	\psi_{+}(\xi^-)~|P\rangle~
\lb{qdq}
\ee
that are recognizable\ct{SPL} as unpolarized, $q_{a}(\xbj)$,
and polarized, $\Delta q_{a}(\xbj)$, 
quark distributions\footnote{
In eq.\rf{qdq} and later
$\psi_+\equiv\frac{1}{2}\ga^{-}\ga^{+}\psi$ is the ``good'' component
of the Dirac field $\psi$, which is the dynamical field in
light-cone quantization\ct{Kogan}.}. One can easily write down the
corresponding distributions $q_{a}(\xbj)$ and $\Delta q_{a}(\xbj)$ 
for a particular flavor $a=u,d,\dots$ complementing $\Ga^q$'s
in eq.\rf{qdq} with a flavor projector 
$P^{(a)}_{ff'}=\delta_{fa}\delta_{f'a}$~.

Similarly we proceed with the gluon structure functions $g(\xbj)$ and 
$\Delta g(\xbj)$. For the first one 
$\Ga^{g\chi}_{\la}=\delta^{\chi}_{\la}$ so that
\be
f_{g}(\xbj)=\frac{i}{4\pi}\xint{\xi^-}\exp{i\xbj P^+\xi^-}
        \langle P|F^{+\la}(0)A_{\la}(\xi^-)~|P\rangle~.
\lb{g}
\ee
Substituting $\exp{i\xbj P^+\xi^-}$ by $(i\xbj P^+)^{-1}
\frac{\partial}{\partial \xi^-}\exp{i\xbj P^+\xi^-}$,
integrating by parts, and remembering that 
$\partial_-A^{LC}_{\la}=F^{+~LC}_{~\la}=-F^{~+LC}_{\la}$ we obtain
\be
f_{g}(\xbj)=\frac{1}{4\pi\xbj P^+}\xint{\xi^-}\exp{i\xbj P^+\xi^-}
  \langle P|F^{+\la}(0)F^{~+}_{\la}(\xi^-)~|P\rangle
  +\left[\left.(\dots)
  \right|^{\xi^-{=}+\infty}_{\xi^-{=}-\infty}\right]~.
\lb{g1}
\ee
The first term on the right hand side of eq.\rf{g1} is 
a standard form of the unpolarized gluon distribution 
$g(\xbj)$\ct{SPL}. It has been argued\ct{BJdG} that
the second, surface, term in eq.\rf{g1} can be dropped. 
We postpone the discussion of the gluon spin dependent distribution
$\Delta g(\xbj)$ until the next section, where it appears in a natural
way.

\section{Angular momentum in deep inelastic scattering}

The application of our formalism for obtaining $\xbj$-distributions 
that describe angular momentum in a hadron is straightforward: 
one should identify 
the corresponding transformations of the quark and gluon fields 
$\tilde{\psi}$ and $\tilde{G}$ and substitute their generators 
$\Ga^q$ and $\Ga^{g}$ into the master 
formulas\rf{fdef}. As mentioned earlier, we are interested in 
transformations that are diagonalized by the 
``parton decomposition''\rf{pd1},\rf{pd2} 
or equivalently that commute with the gauge covariant translations given
by eq.\rf{tfin}. Among three-dimensional rotations these are the
rotations about the $x^3$-axis generated by $J^{12}$ that
transform the fields $\tilde{\psi}(p^+,\tx)$ and
$\tilde{G}_{\la}(p^+,\tx)$ as
\ba{c}
\delta_J\tilde{\psi}=-\epsilon\left[\{
 -\frac{i}{2}\sigma^{12}\}+\{x^{1}\partial_{2}-x^{2}\partial_{1}\}
 \right]\tilde{\psi}~,\\
\delta_J\tilde{G}_{\la}=-\epsilon\left[\{
 -(\delta^{1}_{\la}\delta_{2}^{\chi}-
 \delta^{2}_{\la}\delta_{1}^{\chi})\}+\{x^{1}\partial_{2}-
   x^{2}\partial_{1}\}\right]\tilde{G}_{\la}~.
\lb{qgrot}
\ea
We consider the transformations generated by each of the four terms 
surrounded by braces in eq.\rf{qgrot} independently of the others. 
The first generators in each of these equations are naturally
associated with the field spins whereas the second are the standard 
generators of orbital angular momentum. 
In order obtain to gauge invariant
structure functions we must maintain covariance with respect to the
residual gauge transformations, in which the local gauge transformation 
parameter does not depend on $x^-$.
This can be achieved replacing $\partial_i$ by
the residual gauge covariant derivative
\be
{\cal D}_i\equiv \partial_i-ig{\cal A}_i~.
\lb{calD}
\ee
Similar to the situation at the beginning of the previous section,
the replacement\rf{calD} adds only an overall unphysical gauge 
transformation to the rotation\rf{qgrot} when $\delta_J$ applies
to {\it both} quark and gluon fields with the same 
infinitesimal angle $\epsilon$. 
Also notice that by construction of ${\cal A}_{\la}$ in eq.\rf{calA}
we have ${\cal D}_i=\partial_i$ in any gauge where 
$\lim_{x^-\to\pm\infty}A^{LC}_{i}(x)=0$.

Splitting the transformation\rf{qgrot} 
into the four independent ones as described, we define
\ba{crrl}
\Sigma &: & \delta_{\Sigma}\tilde{\psi}&\equiv
  i\epsilon \sigma^{12}\tilde{\psi}=
  i\epsilon \ga^{0}\ga^{3}\ga^{5}
  \tilde{\psi}~;\\
L_q &: & \delta_{L_q}\tilde{\psi}&\equiv
  -\epsilon(x^{1}{\cal D}_{2}-x^{2}{\cal D}_{1})
  \tilde{\psi}~;\\
\Delta G &: & \delta_{\Delta G}\tilde{G}_{\la}&\equiv
  \epsilon(\delta^{1}_{\la}\delta_{2}^{\chi}-
  \delta^{2}_{\la}\delta_{1}^{\chi})\tilde{G}_{\chi}=
   \epsilon~\e^{+-}{}_{\la}{}^{\chi}\tilde{G}_{\chi}~;\\
L_g &: & \delta_{L_g}\tilde{G}_{\la}&\equiv
 -\epsilon(x^{1}{\cal D}_{2}-x^{2}{\cal D}_{1}) 
 \tilde{G}_{\la}~.
\lb{SLdef}
\ea
The corresponding generators are
\ba{lll}
\Ga_{\Sigma}=\sigma^{12}=\ga^{0}\ga^{3}\ga^{5} &,&~ 
\Ga_{L_q}=(x^{1}i{\cal D}_{2}-x^{2}i{\cal D}_{1})~,\\
(\Ga_{\Delta G})_{\la}{}^{\chi}=-i\e^{+-}{}_{\la}{}^{\chi} &,&~
(\Ga_{L_g})_{\la}{}^{\chi}=(x^{1}i{\cal D}_{2}-
  x^{2}i{\cal D}_{1})\delta^{\chi}_{\la}~,
\ea
where $\e^{\mu\nu\alpha\beta}$ is the antisymmetric Levi-Civita 
tensor with $\e^{0123}\equiv +1$ and $\e^{{+}{-}12}=-1$~.

Following tradition the generator for $\Sigma$ is defined without
the $\frac{1}{2}$ factor that appears in front of $\sigma^{12}$ in
eq.\rf{qgrot}. Substituting these $\Ga$'s into 
eqs.\rf{jg1}--(\ref{jg2}),\rf{fdef}
and remembering that the ${\cal A}_{\la}(\tx)$ term in eq.\rf{jg2} 
can be dropped, we obtain the following structure functions
for the quark and gluon spins and angular momenta:\footnote{
We remind the reader that $\psi_+\equiv\frac{1}{2}\ga^-\ga^+\psi$,
the hadron $|P\rangle$ is polarized along its momentum ${\bf P}$, 
and the fields $\psi$, $A_{\la}$, and $F^{\mu\nu}$ in the structure
functions below are either connected by a Wilson line to make these 
expressions gauge invariant or are evaluated in the gauge $A^{+}{=}~0$, 
where the Wilson line is unity.}
\be
f_{\Sigma}(\xbj)&=&\frac{1}{2\pi\sqrt{2}}\xint{\xi^-}
  \exp{i\xbj P^+\xi^-}\langle P|\psi_{+}^{\dagger}(0)\ga^5
  \psi_{+}(\xi^-)~|P\rangle~; 
\lb{fdq}\\ \nonumber \\
f_{L_q}(\xbj)&=&\frac{\xint{\xi^-}\exp{i\xbj P^+\xi^-}\langle P|
  \int\d^2\xp\psi_{+}^{\dagger}({\xp})(x^{1}i{\cal D}_{2}
  -x^{2}i{\cal D}_{1}) \psi_{+}({\xp}{+}~\xi^-)~|P\rangle}
  {2\pi\sqrt{2}~(\int\d^2\xp)}~;
\lb{flq}\\ \nonumber \\
f_{\Delta G}(\xbj)&=&\frac{1}{4\pi}\xint{\xi^-}
  \exp{i\xbj P^+\xi^-}\langle P|F^{+\la}(0)
  \e^{+-}{}_{\la}{}^{\chi}A_{\chi}(\xi^-)~|P\rangle~;
\lb{fdg}\\ \nonumber \\ 
f_{L_g}(\xbj)&=&\frac{i\xint{\xi^-}\exp{i\xbj P^+\xi^-}\langle P|
  \int\d^2\xp F^{+\la}({\xp})(x^{1}i{\cal D}_{2}
  -x^{2}i{\cal D}_{1})A_{\la}({\xp}{+}~\xi^-)~|P\rangle}
  {4\pi~(\int\d^2\xp)}~.
\lb{flg}\\ \nonumber  
\ee
In these expressions we have translated fields to the origin and 
canceled overall integrals where it is possible. 
One can replace the vector potential $A_{\la}$ in 
eqs.\rf{fdg} and\rf{flg} in favor of the gluon field strength
$F^{~+}_{\la}$ as was done when we derived eq.\rf{g1}.
Dropping the surface terms, we obtain:
(below $\Fd^{\mu\nu}\equiv\frac{1}{2}\e^{\mu\nu\alpha\beta}
F_{\alpha\beta}$)
\be
f_{\Delta G}(\xbj)&=&\frac{i}{4\pi\xbj P^+}\xint{\xi^-}
\exp{i\xbj P^+\xi^-}\langle P|F^{+\la}(0)
\Fd^{~+}_{\la}(\xi^-)~|P\rangle~,~~
\lb{fdf}\\ \nonumber \\
f_{L_g}(\xbj)&=&\frac{\xint{\xi^-}\exp{i\xbj P^+\xi^-}\langle P|
  \int\d^2\xp F^{+\la}({\xp})(x^{1}i{\cal D}_{2}
  -x^{2}i{\cal D}_{1})F^{~+}_{\la}({\xp}{+}~\xi^-)~|P\rangle}
  {4\pi\xbj P^+(\int\d^2\xp)}~.
\lb{flf}\\ \nonumber 
\ee

These are our principal results. Equations\rf{fdq}--(\ref{flf}) 
give a complete set of distribution functions needed to describe 
a hadron's angular momentum. These expressions are automatically
gauge and boost invariant as a consequence of our method.
Our expressions for orbital angular momentum differ from the naive 
gauge invariant form in which $D_i$ appears instead of ${\cal D}_i$.
This distinction is essential because ${\cal D}_i$ depends only on
$\tx$ whereas $D_i$ is also a function of $x^-$, which is impossible to
interpret in a parton model formalism. Our distribution functions 
can be written in an explicitly 
gauge-invariant form with use of the Wilson connection. 
We devote the rest of the section to analyzing
these expressions and comparing them with earlier works. 

In a non-interacting parton model the structure functions
$f_{\Sigma}$, $f_{L_q}$, $f_{\Delta G}$, and $f_{L_g}$ describe
the $\xbj$-distributions of spin and orbital 
angular momentum $x^3$-projections carried by quark and gluon partons 
respectively. This is also true in the interacting theory when
one defines quark and gluon partons as excitations of the fields 
$\tilde{\psi}(p^+,\tx)$ and $\tilde{G}_{\la}(p^+,\tx)$ and 
their spin and orbital momentum as the charges of the 
transformations\rf{SLdef}. We demonstrated that these are the definitions
that reproduce the familiar unpolarized and polarized quark and gluon
structure functions.
According to eq.\rf{1mom},
the first moments of the structure functions
\be
\Ga\equiv \int_{-1}^{1}\d\xbj~ f_{\Ga}(\xbj)~,
\ee
should give the total quark/gluon spin/orbital momentum
in the hadron in the {\it infinite momentum} frame. 
The sum of the first moments obeys the angular momentum sum
rule in eq.\rf{sumrule}, where $1/2$ on the right hand
side is the $x^3$-projection of the nucleon spin and should be 
replaced correspondingly for a hadron with a different spin.
The first moment of $f_{\Sigma}$
\be
\Sigma=\frac{1}{\sqrt{2}P^+} \langle P|
\psi^{\dagger}_{+}(0)\gamma^5\psi_{+}(0)|P\rangle~,
\lb{MS}
\ee
indeed equals the total $x^3$-projection of the quark spin
\be
\Sigma(P') = \frac{\langle P'|\int\d^3 x~\bar\psi(x)\gamma^3
  \gamma^5\psi(x)|P'\rangle}{\langle P'|P'\rangle}=
 \frac{1}{2E'_P}~ 
 \langle P'|\bar\psi(0)\ga^3\ga^5\psi(0)|P'\rangle~
\lb{sigma}
\ee
as $P'^3\to+\infty$. 

The situation for $L_q$, $\Delta G$, and $L_g$ is a little more 
subtle. The first moments of the corresponding structure
functions are 
\be
{L_q}&=&\frac{1}{\sqrt{2}P^+(\int\d^2\xp)} \langle P| \int\d^2\xp 
  \psi^{\dagger}_{+}({\xp})(x^{1}i{\cal D}_{2}-
  x^{2}i{\cal D}_{1})\psi_{+}({\xp})|P\rangle~,\lb{MLq}\\ \nonumber\\
{\Delta G}&=&\frac{1}{2P^+} \langle P|F^{+\la}(0)
  \e^{+-}{}_{\la}{}^{\chi}A_{\chi}(0)~|P\rangle~,\\ \nonumber\\
{L_g}&=&\frac{i}{2P^+(\int\d^2\xp)}
  \langle P| \int\d^2\xp 
  F^{+\la}({\xp})(x^{1}i{\cal D}_{2}
  -x^{2}i{\cal D}_{1})A_{\la}({\xp})~|P\rangle~.\lb{MLg}
\ee
Up to the replacement $\partial_i\leftrightarrow {\cal D}_i$,
the eqs.\rf{MS} and \rf{MLq}--(\ref{MLg})
reproduce the normalized forward matrix elements of the four operators 
in the gauge {\it variant} angular momentum 
decomposition by Jaffe and Manohar\ct{J&M}
\be 
J^3=\int\d^3 x~[~\frac{1}{2}\bar\psi\ga^3\ga^5\psi
  +\psi^{\dagger}(\vec{x}\times i\vec{\partial})^3\psi
  +(\vec{E}\times\vec{A})^3
  -E^k(\vec{x}\times\vec{\partial})^3 A^k]
\lb{J4}
\ee
evaluated in the $A^+{=}~0$ gauge and the infinite momentum frame.
This result is quite welcome in view of previous studies\ct{J&M,JTH,H&S,H&K} 
of angular momentum in deep inelastic scattering. 
This implies that many previous calculations, including
$Q^2$ evolution and asymptotic behavior\ct{JTH,H&S,H&K},
obtained in $A^+{=}~0$ gauge are transferable to our case
with only minor modifications to incorporate $\partial_i\to {\cal D}_i$.

On the other hand, it is also known\ct{Ji} that the total angular 
momentum\footnote{
In the following eq.\rf{efg} $\Theta^{\mu\nu}$ stands for the symmetric 
energy-momentum  Belifante tensor.}
\be
J^k=\frac{1}{2}~\e^{kij}\int\d^3 x(x^{i}\Theta^{0j}-
  x^{j}\Theta^{0i})
\lb{efg}
\ee 
can be decomposed into {\it three} gauge-{\it{}invariant}
vectors:\ct{Ji}
\be
\vec{J}=\int\d^3 x~[~\frac{1}{2}\bar\psi\vec{\ga}\ga^5\psi
  +\psi^{\dagger}(\vec{x}\times i\vec{D})\psi
  +\vec{x}\times(\vec{E}\times\vec{B})]~,
\lb{J3}
\ee
where 
\be
E^k=F^{k0}~,~~B^k=-\frac{1}{2}\e^{klm}F_{lm}
\ee
are the color-electric and color-magnetic field vectors.
The second term in eq.\rf{J3} is an alternative candidate 
for a definition of the quark orbital angular momentum. The third, gluon, 
term does not allow any further separation into local gauge-invariant 
operators that would describe spin and orbital momentum of gluons\footnote{
We were able to give gauge invariant definitions for the gluon spin
and angular momentum {\it projections} only
on the axis $x^3$ defined by the {\it external} momentum $q$.
It is not possible to define gauge invariant tensors of gluon spin and 
orbital momentum independent of a specified direction in space-time.}. 
The $d^3 x$ integrals of each of the three terms in eq.\rf{J3}
are still not covariant under Lorentz boosts so one have to
specify a frame in which they are evaluated.
These quantities are 
in principle measurable and can be related to form-factors of deeply 
virtual Compton scattering\ct{DVCS}. Nevertheless, the decomposition\rf{J3}
is not favorable for deep inelastic scattering experimental conditions 
in the sense that the separate terms in eq.\rf{J3} mix partons with different
$p^+$ and thus can {\it not} be presented in terms of parton densities. 
Of course the greatest shortcoming of the decomposition of eq.\rf{J3}
is its inability to describe the gluon spin spin distribution 
$\Delta g(\xbj)$ which is clearly gauge invariant and can be measured in a 
variety of processes. There is no such a difficulty in the suggested 
approach where $\Delta g(\xbj)=f_{\Delta G}(\xbj)$.

\section{Conclusion}

Typically in (semi)-inclusive deep inelastic scattering the Bjorken 
variable $\xbj$ is fixed, and only forward hadron matrix elements are 
accessible. In order to accommodate these experimental conditions, we
propose to describe a hadron in terms of those observables $\Ga$
that are diagonal in the basis formed by quark and gluon partons.
We then construct gauge invariant $\xbj$-distributions associated 
with such an observable. The gauge covariant definition of parton 
states in fully interacting QCD are chosen so that the parton 
distributions are given by the physical structure functions 
$q(\xbj)$ and $g(\xbj)$. To satisfy this requirement,
quark and gluon partons must be eigenstates  of the 
generators of covariant translations $T^q_-$ and $T^g_-$, eq.\rf{tfin},
and the $\Ga$ distributions are given by the formulas
\be
f_{\Ga^{q}}(\xbj)=\frac{\langle P|
\int\d^2\xp\xint{x^-}\xint{\xi^-}~\exp{i\xbj P^+\xi^-}
\bar\psi(x)\ga^{+}\Ga^{q}\psi(x+\xi^-)~|P\rangle}
{4\pi(\int\d^2\xp\xint{x^-})}~
\lb{fmq}
\ee
and
\be
f_{\Ga^{g}}(\xbj)=\frac{\langle P|
\int\d^2\xp\xint{x^-}\xint{\xi^-}~\exp{i\xbj P^+\xi^-}
F^{+\la}(x)\Ga^{g\chi}_{\la} F^{~+}_{\chi}(x+\xi^-)~|P\rangle}
{4\pi\xbj P^+(\int\d^2\xp\xint{x^-})}~
\lb{fmg2}
\ee
in $A^+{=}~0$ gauge. They may be put in a 
gauge-invariant form by trivial use of Wilson lines. In these formulas 
the generators $\Ga^{q}$ and $\Ga^{g}$ transform under the 
residual gauge group as given by eq.\rf{Garest} and are uniquely 
related to the original quantity $\Ga$.

A hadron angular momentum is completely  described by the four scale 
dependent
$\xbj$-distributions given by eqs.\rf{fdq}-(\ref{flq}),\rf{fdf}-(\ref{flf}).
Two of them, $f_{\Sigma}$ and $f_{\Delta G}$, coincide with the polarized
quark and gluon structure functions $\Delta q$ and $\Delta g$.
The other two are naturally regarded as the $\xbj$ distributions 
of quark and gluon orbital angular momentum. 
They are well defined physical objects and are gauge invariant by  
means of the residual gauge covariant derivative
${\cal D}_i=\partial_i-ig{\cal A}_i$, where ${\cal A}$ is given by 
eq.\rf{calA}. They are observable in principle 
and can be calculated in models or within lattice QCD.
However, it remains an open question whether the $\xbj$-distributions of 
quark and gluon orbital angular momentum which we have defined 
can be measured in a practical experimental process.

\section*{Acknowledgments}

We are grateful to J.~Goldstone, D.~Dolgov, J.~Negele, and
A.~Pochinsky for helpful discussions. R.L.J. would also like to thank
the RIKEN-BNL Research Center for partial support.


\begin{references}

\bibitem{EMC}
J. Ashman et al.,
\Journal{\PLB}{206}{364}{1988};\\
J. Ashman et al.,
\Journal{\NPB}{328}{1}{1988}.

\bibitem{Ratcl}
P.~G.~Ratcliffe,
\Journal{\PLB}{192}{180}{1987}.

\bibitem{C&S}
J.~C.~Collins and D.~E.~Soper,
\Journal{\NPB}{194}{445}{1982}.

\bibitem{dG}
A. V. Manohar,
\Journal{\PLB}{255}{579}{1991};\\
I. I. Balitskii and V. M. Braun, 
\Journal{\PLB}{267}{405}{1991}.

\bibitem{BJdG}
R.~L.~Jaffe, 
\Journal{\PLB}{365}{359}{1996}.

\bibitem{J&M}
R.~L.~Jaffe and A.~Manohar,
\Journal{\NPB}{337}{509}{1990}.

\bibitem{newJi}
After the completion of this paper we became aware of a recent work by
P.~Hoodbhoy, X.~Ji, and W.~Lu that addresses some similar issues: hep-ph/9804337. 

\bibitem{JTH}
X.~Ji, J.~Tang, and P.~Hoodbhoy,
\Journal{\PRL}{76}{740}{1996}.

\bibitem{H&S}
P.~H\"{a}gler and A.~Sch\"{a}fer, hep-ph/9802362.

\bibitem{H&K}
A. Harindranath, R. Kundu, hep-ph/9802406.

\bibitem{Ji}
X.~Ji,
\Journal{\PRL}{78}{610}{1997}.

\bibitem{Kogan}
J.~B.~Kogut and D.~E.~Soper,
\Journal{\PRD}{1}{2901}{1970}.

\bibitem{UVD_C_p447}
J. C. Collins, \Journal{\PRD}{21}{2962}{1980}.

\bibitem{trans}
J. Ralston and D. E. Soper, 
\Journal{\NPB}{152}{109}{1979};\\
J. L. Cortes, B. Pire, and J. P. Ralston,
\Journal{\ZPC}{55}{409}{1992};\\
R. L. Jaffe and X. Ji, 
\Journal{\PRL}{67}{552}{1991};
\Journal{\NPB}{375}{527}{1992}.

\bibitem{SPL}
See, for example, R. L. Jaffe,
{\it Spin, Twist, and Hadron Structure in Deep Inelastic Processes},
Lectures given at Ettore Majorana International School of Nucleon Structure
(1995): hep-ph/9602236.

\bibitem{DVCS}
X. Ji, \Journal{\PRD}{55}{7114}{1997}.

\end{references}
\end{document}